\documentstyle[multicol,aps,epsf]{revtex}
\begin{document}
\title{Organization of Growing Random Networks}
\author{P.~L.~Krapivsky and S.~Redner}
\address{Center for BioDynamics, Center for Polymer Studies, 
and Department of Physics, Boston University, Boston, MA, 02215}

\maketitle
\begin{abstract}
  
  The organizational development of growing random networks is investigated.
  These growing networks are built by adding nodes successively and linking
  each to an earlier node of degree $k$ with attachment probability $A_k$.
  When $A_k$ grows slower than linearly with $k$, the number of nodes with
  $k$ links, $N_k(t)$, decays faster than a power law in $k$, while for $A_k$
  growing faster than linearly in $k$, a single node emerges which connects
  to nearly all other nodes.  When $A_k$ is asymptotically linear,
  $N_k(t)\sim tk^{-\nu}$, with $\nu$ dependent on details of the attachment
  probability, but in the range $2<\nu<\infty$.  The combined age and degree
  distribution of nodes shows that old nodes typically have a large degree.
  There is also a significant correlation in the degrees of neighboring
  nodes, so that nodes of similar degree are more likely to be connected.
  The size distributions of the in-components and out-components of the
  network with respect to a given node -- namely, its ``descendants'' and
  ``ancestors'' -- are also determined.  The in-component exhibits a robust
  $s^{-2}$ power-law tail, where $s$ is the component size.  The out
  component has a typical size of order $\ln t$ and it provides basic
  insights about the genealogy of the network.

\end{abstract}
\begin{multicols}{2}
  
\section{Introduction}

Networks of many interacting units play an important role in epidemiology,
ecology, gene regulation, neural networks, and many other
fields\cite{food,kauf,neural}.  In many studies of these networks, the number
of nodes is considered to be fixed and the presence of a link between two
nodes is treated as a random event independent of the other links.  These
assumptions lead naturally to random graph models\cite{bol,jan}.  While these
models have rich behavior and considerable utility, they are not necessarily
appropriate for describing {\em growing} networks, where the addition of
nodes and links may depend on the local features of the network where the
growth event is taking place.

Typical examples of such growing networks include transportation or
electrical distribution systems, where growth occurs in response to
population-driven demands.  Two currently appealing examples are the
distribution of scientific citations and the structure of the world-wide web.
For both these examples there is now considerable data available, in spite of
the very rapid growth of these systems.  In the former case, one may consider
papers to be the nodes of a graph and citations as the links.  The structure
of the resulting ``citation graph'' was originally studied by Lotka in
1926\cite{lotka}, and then by many others
\cite{shock,gar,intr,gil,sor,redner,new}.  The basic feature of this citation
distribution is that it appears to have a relatively steep power-law tail;
thus most papers are minimally cited while highly-cited papers are rare.

Similarly, in the web graph, much structural data has recently been obtained
\cite{kum1,klein,kum2,hub1,hub2,fal,b1,matta} which suggest that the number
of nodes with $k$ links has a power-law tail, with an exponent that is
somewhat larger than 2.  This power-law tail again corresponds to the basic
fact that most nodes of the web graph are unimportant, while a relatively
small number of nodes garner a large fraction of ``hits''.  Due to the
qualitative similarities between the citation and web graphs, insights
developed in the field of bibliometrics\cite{intr} have been applied to help
understand the structure of the web\cite{larson}.

Because of the dynamic nature of the citation and web graphs, it is not
surprising that their topologies at any fixed time are very different from
classical random graphs.  In distinction to the power-law degree
distributions of the citation and web graphs, random graphs have a Poisson
node degree distribution.  Here node degree is defined as the number of links
at a node.  To overcome the shortcomings of random graphs in describing the
dynamic natures of these systems, both ``small-world'' networks
\cite{small,watts} and growing random network models\cite{b1,b2,j1,j2,krl}
have been recently introduced.  The former are aimed at understanding the
relatively small diameter of large graphs of socially interacting units,
while the latter seek to understand the growth dynamics.

In this paper, we provide a comprehensive quantitative description of a
simple {\em growing network} (GN) model.  Our results are based on the
analysis of the rate equations for the densities of nodes of a given degree.
This approach bears many similarities to the rate equations for the kinetics
of aggregation.  The rate equations for the evolution of growing networks are
relatively simple and the results that emerge are comprehensive.  Thus it
appears that the rate equation method is better suited for probing the
structure of growing random networks compared to the classical approaches for
analyzing random graphs, such as probabilistic\cite{bol} or generating
function\cite{jan} techniques.  The rate equation approach also has the
advantage that it can be adapted to other evolving graph systems, including
networks with addition and deletion of nodes and links, as well as networks
with link re-wiring.

\begin{figure}
  \narrowtext \hskip 0.6in \epsfxsize=2.3in 
\epsfbox{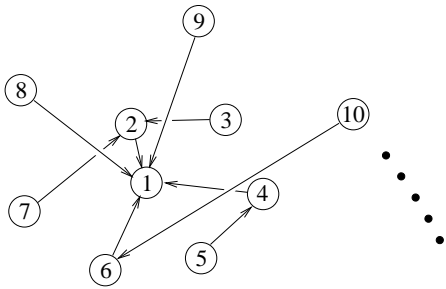} \vskip 0.1in
\caption{Schematic illustration of the evolution of the growing random 
  network.  Nodes are added sequentially and a single link joins the new node
  to an earlier node.  In this example, node 1 has degree 5, node 2 has
  degree 3, nodes 4 and 6 have degree 2, and all the remaining nodes have
  degree 1.  Also note that node 1 is the ``ancestor'' of 6, while 10 is the
  descendant of 6.
\label{network}}
\end{figure}

We will specifically investigate two types of models: (a) the GN in which
nodes are added one at a time and a link is established with a pre-existing
node according to an attachment probability $A_k$ which depends only on the
degree of the target node (Fig.~\ref{network}), and (b) the GN with
re-direction (GNR), in which the newly-created link can be re-directed to the
``ancestor'' node of the original target node.  An important feature of these
models is that the links are {\em directed\/} and the resulting graphs have a
simple tree-like topology.  The motivation for the GNR model is that this
re-direction process roughly mimics how we might (lazily) construct the
references to this paper.  In addition to papers that we peruse and cite
directly, we are also likely to incorporate some of the references within
these papers as part of our reference list.  A related ``copying'' process
also affects the organization of the web\cite{klein}.

One of our primary results is that for asymptotically linear attachment
kernels, $A_k\sim k$ as $k\to\infty$, the degree distribution of the GN has
a power-law form $N_k(t)\sim tk^{-\nu}$, with $\nu$ tunable in the range
$2<\nu<\infty$.  By choosing the control parameters of our model in a
plausible manner it is then easy to reproduce the quantitative observations
about the node degree distribution of the web graph.

In Sec.~II, we define the GN and GNR models precisely and then determine
their node degree distributions in Sec.~III by the rate equation approach.
Different distributions arise in the GN model which depend on the asymptotic
behavior of the attachment probability as a function of node degree.  In
Sec.~IV, we investigate the joint age-degree distribution and find (not
surprisingly) that ``old'' nodes are typically more highly connected.  In
Sec.~V, we study the correlations which develop between the degrees of
connected nodes as the network grows.  In Sec.~VI, we study a more global
measure of the network, namely, the size distributions of the
``in-component'' and ``out-component''.  With respect to a given node {\bf
  x}, the in-component is the set of nodes which can reach node {\bf x} via a
directed path of links.  Conversely, the out-component is the set nodes which
can be reached from node {\bf x} via a directed path.  The former exhibits a
robust power-law size distribution which appears to be independent of the
attachment probability.  The latter distribution predicts a network
``diameter'' which grows as $\ln t$ and also provides basic insights about
the genealogy of the network.  We conclude in Section VII.

\section{The Models}

\subsection{Growing Network (GN)} 

In the GN, we introduce a new node at each time step and link it to one of
the earlier nodes in the network (Fig.~\ref{network}).  This leads to a
network which has a topology of a (directed) tree graph.  In terms of
citations, we may interpret the nodes as publications, and the directed link
from one paper to another as a citation to the earlier publication.  In terms
of the web graph, nodes are web pages and the directed links are hyperlinks.
We will refer to the node to which the link is directed as the {\em ancestor}
of the current node.

As the network grows, a degree distribution $N_k(t)$, defined as the average
number of nodes with $k$ links ($k-1$ incoming and 1 outgoing) builds up.
The initial node is unique as it does not have an outgoing link.  The basic
ingredient which determines the structure of the network is the {\em
  attachment kernel\/} $A_k$, defined as the probability that the
newly-introduced node links to an existing node which already has $k$ links.
On general grounds, this attachment kernel should be a non-decreasing
function of $k$, and natural scenarios are attachment kernels with a power
law dependence on $k$.  For the linear kernel, the GN reduces to the ``scale
free'' model introduced by Barab\'asi and Albert\cite{b1} and further
investigated in \cite{b2,j1,j2}.

The general homogeneous model, $A_k=k^\gamma$ with $\gamma\geq 0$, was
investigated in\cite{krl} where it was found that the degree distribution
$N_k(t)$ crucially depends on the value of $\gamma$.  For $\gamma<1$, the
linking probability grows weakly with node ``popularity'' and $N_k(t)$
decreases as a stretched exponential in $k$ for any $t$.  The complementary
case of $\gamma>1$ leads to phenomenon akin to gelation\cite{agg} in which a
single ``gel'' node links to nearly every other node.  For $\gamma>2$,
this phenomenon is so extreme that the number of links between other nodes is
finite in an infinite graph.  We shall show that these results also apply for
the more general situation where $A_k\sim k^\gamma$ as $k\to\infty$ in
addition to the strictly homogeneous situation where $A_k=k^\gamma$.

The borderline case of an asymptotically linear attachment kernel, $A_k\sim
k$, is particularly intriguing as it leads to $N_k\sim k^{-\nu}$, with the
exponent $\nu$ tunable to {\em any} value larger than 2 depending on finer
details of the attachment kernel.  In particular, the strictly linear kernel,
$A_k=k$, leads to $\nu=3$.  However, by changing the value of a single
attachment probability, for example $A_1=\alpha$ and $A_k=k$ for $k>2$, any
value of $\nu>2$ is possible.  This sensitivity of asymptotic behavior on
microscopic details indicates that the case of attachment index $\gamma=1$ is
marginal.  A related phenomenon occurs in constant-kernel aggregation, where
the asymptotic kinetics is sensitively dependent on the actual values of the
reaction rate\cite{lr,cl}.

\subsection{Growing Network with Re-direction (GNR)}

\begin{figure}
  \narrowtext \hskip 0.6in \epsfxsize=1.8in
\epsfbox{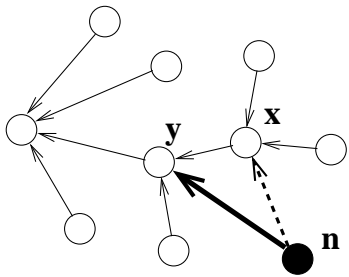} \vskip 0.1in
\caption{
  Illustration of the basic processes in the GNR model.  The new node 
  (solid) selects a target node {\bf x}.  With probability $1-r$ a link is
  established to this target node (dashed arrow), while with probability $r$
  the link is established with the ancestor of {\bf x} (thick solid arrow).
\label{R}}
\end{figure}

The GN is built by simultaneous node and link addition and disregards other
elemental processes which can occur in the development of large networks.  In
the context of the web, these include node and link deletion (for out-of-date
websites), link re-wiring, the tendency of a new node to connect to nearby
nodes, and the copying of links from existing nodes to new nodes.  The GNR
model incorporates a simple form of link re-wiring into the GN model.  At
each time step, a new node {\bf n} is added and an earlier node {\bf x} is
selected {\em uniformly} as a possible ``target'' for attachment.  With
probability $1-r$, the link from {\bf n} to {\bf x} is created; in this case,
the evolution is the same as in the GN.  However, with probability $r$, the
link is {\em re-directed\/} to the ancestor node {\bf y} of node {\bf x}
(Fig.~\ref{R}).

A model of this spirit was recently mentioned in the context of the web
development\cite{klein}.  A related model was also proposed long ago by
Simon\cite{simon,simon2} to describe the word frequencies of English text.
The Simon model gives a power-law frequency distribution whose exponent is
tunable in manner which closely mirrors the behavior in the GNR model.  The
Simon model was also recently applied to explain power-law distributions in
the frequency of family names\cite{zan}.

While at first sight the GNR model appears complicated, we shall see that its
characteristics can be obtained in a simple fashion.  Another very helpful
and surprising property of the GNR with a uniform initial attachment
probability is that it is equivalent to the GN with a shifted linear
attachment kernel and {\em no} re-direction.  We shall exploit this
equivalence extensively in the following.  Nevertheless, we consider the GNR
separately, as in many cases the rate equations for the GNR with a {\em
  uniform} attachment kernel is simpler to appreciate than the rate equations
for the GN with a shifted linear attachment kernel.

\section{The degree distribution}

\subsection{GN Model}

We now study the evolution of the degree distribution of the GN model.  The
rate equations for $N_k(t)$ are
\begin{equation}
\label{Nk}
{d N_k\over dt}=A^{-1}
\left[A_{k-1} N_{k-1}-A_k N_k\right]+\delta_{k1}.
\end{equation}
The first term on the right-hand side of Eq.~(\ref{Nk}) accounts for the
process in which a node with $k-1$ links is connected to the new node,
leading to a gain in the number of nodes with $k$ links.  This happens with
probability $A_{k-1}/A$, where $A(t)=\sum_{j\geq 1} A_jN_j(t)$ is the
appropriate normalization factor.  A corresponding role is played by the
second (loss) term on the right-hand side of Eq.~(\ref{Nk}).  Notice that the
overall amplitude in $A_k$ is irrelevant, since it appears in both the
numerator and denominator of Eq.~(\ref{Nk}), and can be chosen arbitrarily.
The last term on the right-hand side of Eq.~(\ref{Nk}) accounts for the
continuous introduction of new nodes with no incoming links.  We also set
$N_0\equiv 0$, so that Eq.~(\ref{Nk}) applies for all $k\geq 1$.

It is worth noting that at a fundamental level, Eqs.~(\ref{Nk}) describe the
symbolic reaction $[k]\to[k+1]$.  Many other reactions, such as the
Becker-D\"oring theory of nucleation\cite{bd}, additive
polymerization\cite{he}, hydrolysis\cite{mg}, catalysis and submonolayer
epitaxial growth\cite{epitaxy}, fit into this scheme.  However, there is one
important difference in that we consider strictly a single connected cluster
(the growing network), while in the context of aggregation-like processes,
one generally deals with a collection of clusters.  The effect of having more
than one cluster in the framework of growing networks is currently under
investigation\cite{krrp}.

We start by solving the equations for the low-order moments of the degree
distribution, which are defined by $M_n(t)=\sum_{j\geq 1} j^n N_j(t)$.
Summing Eqs.~(\ref{Nk}) over all $k$ gives the rate equation for the total
number of nodes, $\dot M_0=1$, whose solution is $M_0(t)= M_0(0)+t$.  The
first moment (the total number of link endpoints) obeys $\dot M_1=2$, which
gives $M_1(t)= M_1(0)+2t$.  The first two moments are therefore {\em
  independent\/} of the attachment kernel $A_k$, while higher moments and the
degree distribution itself do depend on the kernel $A_k$.

To develop an appreciation for the types of behavior that can occur, consider
the linear kernel $A_k=k$ for which $A(t)$ coincides with $M_1(t)$.  In this
case, we can solve Eqs.~(\ref{Nk}) for an arbitrary initial condition.
However, since the long-time behavior is most interesting we limit ourselves
to the asymptotic regime ($t\to\infty$) where the initial condition is
irrelevant.  Using therefore $M_1=2t$, we solve the first few of
Eqs.~(\ref{Nk}) and obtain $N_1=2t/3$, $N_2=t/6$, {\it etc.}, which implies
that the $N_k$ grow linearly with time.  Accordingly, we substitute
$N_k(t)=t\,n_k$ in Eqs.~(\ref{Nk}) to yield the simple recursion relation
$n_k=n_{k-1} (k-1)/(k+2)$.  Solving for $n_k$ gives
\begin{equation}
\label{nk1}
n_k={4\over k(k+1)(k+2)}. 
\end{equation}
In the context of discrete functions defined on the positive integers, this
distribution is algebraic over the entire range of $k$.  Indeed, as explained
in Ref.\cite{knuth}, the proper analog of the {\em continuous} power-law
function $f(x)=x^{-\lambda}$ is the {\em discrete} function
$f_k=\Gamma(k)/\Gamma(k+\lambda)$, where $\Gamma$ is the Euler gamma
function.  Rewriting Eq.~(\ref{nk1}) as $n_k=4\Gamma(k)/\Gamma(k+3)$, we see
that $n_k$ is indeed algebraic over the entire range $k\geq 1$.

Returning to more general attachment kernels, let us assume that the degree
distribution and $A(t)$ both grow linearly with time.  We anticipate that
this hypothesis will hold for attachment kernels which do not grow faster
than linearly with $k$.  By substituting $N_k(t)=t\,n_k$ and $A(t)=\mu t$
into Eqs.~(\ref{Nk}) we obtain the recursion relation $n_k=n_{k-1}
A_{k-1}/(\mu+A_k)$ and $n_1=\mu/(\mu+A_1)$.  Solving for $n_k$, we obtain
\begin{equation}
\label{Nkgen}
n_k={\mu\over A_k}\prod_{j=1}^{k}
\left(1+{\mu\over A_j}\right)^{-1}.
\end{equation}
To complete the solution, we need to find the amplitude $\mu$.
Combining the definition $\mu=\sum_{j\geq 1}A_jn_j$ and
Eq.~(\ref{Nkgen}), we obtain the implicit relation
\begin{equation}
\label{mugen}
\sum_{k=1}^\infty \prod_{j=1}^{k}
\left(1+{\mu\over A_j}\right)^{-1}=1.
\end{equation}
Thus the amplitude $\mu$ always depends on the entire attachment kernel.  On
the other hand, we shall show that the degree distribution exhibits a robust
behavior which depends only on gross features of the attachment kernel, as
long as $A_k$ grows {\em slower} than linearly.  The case where $A_k$ is
asymptotically linear is perhaps the most intriguing as the degree
distribution has a power-law behavior whose exponent depends on microscopic
details of the dependence of $A_k$ on $k$.  When $A_k$ grows {\em faster}
than linearly, drastically different gelation-like behavior arises.  It is
again worth noting that these three regimes of kinetic behavior also arise in
the solutions to the rate equations for additive polymerization processes,
with the different regimes arising when the attachment exponent $\gamma$ is
smaller than, larger than, or equal to one\cite{addition}.  

We now separately describe these three cases.

\subsubsection{Sub-linear kernels}

Consider sub-linear kernels which are {\em asymptotically homogeneous}, that
is, $A_k\sim k^\gamma$, with $0<\gamma<1$.  Substituting this asymptotics
into Eq.~(\ref{Nkgen}), writing the product as the exponential of a sum,
converting the sum to an integral, and performing this integral, we obtain
\begin{eqnarray}
\label{cases} 
n_k\sim\cases{
k^{-\gamma}\exp
\left[-\mu\left({{k^{1-\gamma}-2^{1-\gamma}}\over 1-\gamma}\right)\right]
&${1\over 2}<\gamma<1$,\cr
&\cr
k^{\mu^2-1\over 2}\exp\left[-2\mu\,\sqrt{k} \right] 
&$\gamma={1\over 2}$,\cr
&\cr
k^{-\gamma}\exp\left[-\mu\, {k^{1-\gamma}\over 1-\gamma}
+{\mu^2\over 2}\, {k^{1-2\gamma}\over 1-2\gamma}\right]
&${1\over 3}<\gamma<{1\over 2}$,}
\end{eqnarray}
{\it etc.}.  The pattern given in Eq.~(\ref{cases}) continues {\it ad
  infinitum}: Whenever $\gamma$ decreases below $1/m$, with $m$ a positive
integer, an additional term in the exponential arises from the now relevant
contribution of the next higher-order term in the expansion of the product in
Eq.~(\ref{Nkgen}).

\begin{figure}
  \narrowtext \epsfxsize=2.5in \hskip 0.3in
\epsfbox{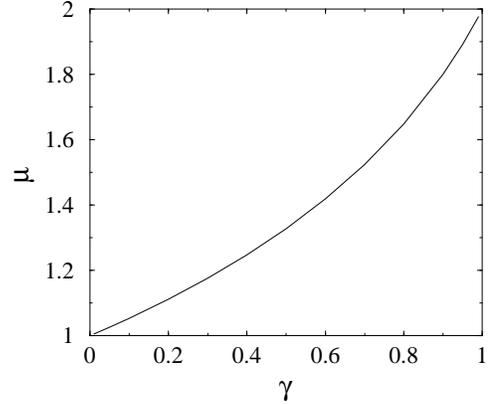} \vskip 0.10in
\caption{The amplitude $\mu$ in $M_\gamma(t)=\mu t$ versus $\gamma$.
\label{mu-vs-gamma}}
\end{figure}

To complete the solution, we require the amplitude $\mu$.  We have been
unable to find an explicit expression for $\mu$, even if the attachment
kernel is strictly homogeneous, $A_k=k^\gamma$, as it requires solving
Eq.~(\ref{mugen}).  However, this relation can be easily evaluated
numerically and it shows that $\mu(\gamma)$ varies smoothly between 1 and 2
as $\gamma$ increases from 0 to 1 (Fig.~\ref{mu-vs-gamma}).  These two limits
correspond to the known limiting behaviors for $M_0$ and $M_1$.

More detailed results can be obtained for the limiting solvable cases of
$A_k={\rm const.}$ and $A_k=k$.  In these limits, $\mu=1$ and $\mu=2$,
respectively, and the corresponding degree distributions are given by
$n_k=2^{-k}$ and by Eq.~(\ref{nk1}).  The former can be easily obtained by
following exactly the same steps as those used to solve the network with the
linear kernel.  We can then apply perturbation theory to find the respective
limiting behaviors of $\mu(\gamma)$ for $\gamma$ close to 0 or 1,
\begin{eqnarray*}
\mu&=&1+B_0\gamma+{\cal O}\left(\gamma^2\right),\\
\mu&=&2-B_1(1-\gamma)+{\cal O}\left((1-\gamma)^2\right),
\end{eqnarray*}
with
\begin{eqnarray*}
B_0&=&\sum_{j=1}^\infty {\ln j\over 2^j}=0.5078 \ldots,\\
B_1&=&4\sum_{j=1}^\infty {\ln j\over (j+1)(j+2)}=2.407 \ldots.
\end{eqnarray*}

\subsubsection{Linear kernels}

Consider now {\em asymptotically} linear attachment kernels, $A_k\sim k$ as
$k\to\infty$.  As already mentioned, we can always choose the amplitude in
the asymptotic relation equal to one, as attachment kernels which differ by a
multiplicative factor give identical behavior.  For the asymptotically linear
kernel, expanding the product in Eq.~(\ref{Nkgen}) and following step-by-step
the approach that led to Eq.~(\ref{cases}) now gives the power-law asymptotic
behavior
\begin{equation}
\label{nula}
n_k\sim k^{-\nu},\qquad {\rm with} \quad \nu=1+\mu. 
\end{equation}
An important feature of this result is that the exponent $\nu$ can be tuned
to {\em any} value larger than 2.  This lower bound immediately follows from
the fact that the sum $\mu=\sum_j A_jn_j\sim \sum_j jn_j$ must converge and
this, in turn, requires that $\nu$ must be larger than 2.

As an explicit example, consider the attachment kernel 
$A_k=k$ for $k\geq 2$, while $A_1\equiv \alpha$ is an arbitrary
positive number.  Now it is convenient to separately treat $A_1$ and $A_k$
for $k\geq 2$ in Eq.~(\ref{mugen}) to recast it as
\begin{equation}
\label{mu2}
\mu=A_1\sum_{k=2}^\infty \prod_{j=2}^{k}
\left(1+{\mu\over A_j}\right)^{-1}.
\end{equation}
The right-hand side of Eq.~(\ref{mu2}) can be simply expressed as the ratio
of Euler gamma functions to yield
\begin{equation}
\label{muv}
\mu=\alpha\sum_{k=2}^\infty \Gamma(2+\mu)\,
{\Gamma(1+k)\over \Gamma(1+\mu+k)}.
\end{equation}
This sum can be evaluated by employing the identity\cite{knuth}
\begin{equation}
\label{ident}
\sum_{k=2}^\infty {\Gamma(a+k)\over \Gamma(b+k)}=
{\Gamma(a+2)\over (b-a-1)\Gamma(b+1)},
\end{equation}
so that Eq.~(\ref{muv}) reduces to $\mu(\mu-1)=2\alpha$, with solution
$\mu=(1+\sqrt{1+8\alpha})/2$.  Thus the exponent $\nu=1+\mu$ is
\begin{equation}
\label{nu}
\nu={3+\sqrt{1+8\alpha}\over 2}.
\end{equation}

Furthermore, following the steps that lead to Eq.~(\ref{Nkgen}), the degree
distribution for the GN with the attachment kernel $A_1=\alpha$ and $A_k=k$
for $k\geq 2$ is
\begin{equation}
\label{nkv}
n_1={\mu\over \mu+\alpha}, \quad 
n_k={\mu \alpha\over \mu+\alpha}\,
{\Gamma(2+\mu)\,\Gamma(k)\over \Gamma(1+\mu+k)}.
\end{equation} 
Notice that for $0<\alpha<1$, the exponent lies in the range $2<\nu<3$; in
particular, $\nu=2+2\alpha-4\alpha^2+\ldots$ as $\alpha\to 0$.  When
$\alpha=1$, we recover the connectively distribution of Eq.~(\ref{nk1}).  For
$\alpha>1$, we have $\nu>3$; in particular, $\nu\to \sqrt{2\alpha}$ as
$\alpha\to \infty$.

The GN is also solvable when $A_k=k+w$.  This shifted linear kernel can be
motivated naturally by explicitly keeping track of the directionality of the
links.  In particular, the node degree for an undirected graph generalizes to
the in-degree and out-degree for a directed graph.  These are just the number
of incoming and outgoing links at a node, respectively.  Thus, the node
degree $k$ in a directed graph is the sum of the in-degree $i$ and out-degree
$j$.  The most general linear attachment kernel for a directed graph is
therefore of the form $A_{ij}=ai+bj$.  The GN corresponds to the case where
the out-degree of any node equals one; thus $j=1$ and $k=i+1$.  Hence the
general linear attachment kernel reduces to $A_k=a(k-1)+b$.  Since, as
mentioned above, the overall scale factor in the kernel is irrelevant, we can
re-write $A_k$ as the shifted linear kernel $A_k=k+w$, with $w=-1+b/a$, so
that it can vary over the range $-1<w<\infty$.

We can now easily determine the degree distribution for the shifted linear
attachment kernel.  First we note that $A(t)=\sum_jA_jN_j=M_1(t)+wM_0(t)$.
Then using the basic results $A=\mu t$, $M_0=t$ and $M_1=2t$, we have
$\mu=2+w$ and thence $\nu=3+w$, according to Eq.~(\ref{nula}).  Furthermore,
from Eq.~(\ref{Nkgen}) we easily determine the entire degree distribution to
be
\begin{equation}
\label{nkw}
n_k=(2+w)\,{\Gamma(3+2w)\over \Gamma(1+w)}\,
{\Gamma(k+w)\over \Gamma(k+3+2w)}.
\end{equation}
In a similar vein, we can solve the GN with an arbitrary piecewise linear
attachment kernel.  In all these cases, the exponent $\nu$ can be tuned to
any value larger than 2, and for sufficiently large degree $n_k$ can be
expressed as the ratio of gamma functions, {\it i.\ e.}, the degree
distribution is a purely (discrete) algebraic function.

\subsubsection{Super-linear kernels}

For the super-linear homogeneous attachment kernels, $A_k=k^\gamma$ with
$\gamma>1$, we now show that a ``winner take all'' phenomenon arises, namely,
there emerges a single dominant ``gel'' node which is linked to almost every
other node.  A particularly singular behavior occurs for $\gamma>2$, where
there is a non-zero probability that the initial node is connected to {\em
  every} other node of the network.

Let us first determine the probability that the initial node connects to all
other nodes.  It is convenient to consider a discrete time version of the GN
in which one node is introduced at each elemental step which always links to
the initial node.  After $N$ steps, the probability that the new node will
link to the initial node is $N^\gamma/(N+N^\gamma)$.  This probability that
this connectivity pattern continues indefinitely is
\begin{equation}
\label{prob}
{\cal P}=\prod_{N=1}^\infty {1\over 1+N^{1-\gamma}}.
\end{equation}
Clearly, ${\cal P}=0$ when $\gamma\leq 2$ but ${\cal P}>0$ when $\gamma>2$.
Thus for $\gamma>2$ there is a non-zero probability that the initial node
connects to all other nodes.  

To determine the behavior for general $\gamma>1$, we first need the
asymptotic time dependence of $M_\gamma$.  To this end, it is useful to
consider the discretized version of the master equations Eq.~(\ref{Nk}),
where the time $t$ is limited to integer values.  Then $N_k(t)=0$ whenever
$k>t$ and the rate equation for $N_k(k)$ immediately leads to
\begin{eqnarray}
N_k(k) &=& \frac{(k-1)^\gamma N_{k-1}(k-1)}{M_\gamma(k-1)}\nonumber \\
       &=& N_2(2)\prod_{j=2}^{k-1}\frac{j^\gamma}{M_\gamma(j)}.
\label{eq:Nkk}
\end{eqnarray}
{}From this, and the obvious fact that $N_k(k)$ must be less than unity, it
follows that $M_\gamma(t)$ cannot grow more slowly than $t^\gamma$.  On the
other hand, $M_\gamma(t)$ cannot grow faster than $t^\gamma$, as follows
from the estimate
\begin{eqnarray}
M_\gamma(t)&=&\sum_{k=1}^t k^\gamma N_k(t)\nonumber \\
  &\leq& t^{\gamma-1}\sum_{k=1}^t k N_k(t)=t^{\gamma-1}M_1(t)
\end{eqnarray}
Thus $M_\gamma\propto t^\gamma$.  In fact, the amplitude of $t^\gamma$ is
unity as we will derive self-consistently after solving for the $N_k$'s.

We now use $M_\gamma\sim t^\gamma$, with $\gamma>1$, in the rate equations to
solve recursively for each $N_k$.  Starting with the equation $\dot
N_1=1-N_1/M_\gamma$, we see that the second term on the right-hand side is
sub-dominant.  Thus by neglecting this term we obtain $N_1=t$.  Similarly,
$\dot N_2=(N_1-2^\gamma N_2)/M_\gamma\sim N_1/M_\gamma$ gives $N_2\sim
t^{2-\gamma}/(2-\gamma)$.  Continuing this same line of reasoning for each
successive rate equation gives the leading behavior of $N_k$,
\begin{equation}
\label{Nkgg}
N_k(t) = J_k t^{k-(k-1)\gamma}\quad {\rm for}\quad k\geq 1,
\end{equation}
with $J_k=\prod_{j=1}^{k-1}j^\gamma/[1+j(1-\gamma)]$.  This pattern of
behavior for $N_k$ continues as long as its exponent $k-(k-1)\gamma$ remains
positive, or $k<\gamma/(\gamma-1)$.  The full temporal behavior of the
$N_k(t)$ may be determined straightforwardly by keeping the next correction
terms in the rate equations.  For example,
$N_1(t)=t-t^{2-\gamma}/(2-\gamma)+\ldots$.  

For $k>\gamma/(\gamma-1)$, each $N_k$ has a finite limiting value in the
long-time limit.  Since the total number of connections equals $2t$, and
$t$ of them are associated with $N_1$, the remaining $t$ links must all
connect to a single node which has $t$ connections (up to corrections
which grow no faster than sub-linearly with time).  Consequently the
amplitude of $M_\gamma$ equals unity, as argued above.

Therefore for super-linear kernels, the GN undergoes an infinite sequence
of connectivity transitions as a function of $\gamma$.  For $\gamma>2$ all
but a finite number of nodes are linked to the ``gel'' node which has the
rest of the links of the network.  This is the ``winner take all'' situation.
For $3/2<\gamma<2$, the number of nodes with two links grows as
$t^{2-\gamma}$, while the number of nodes with more than two links is again
finite.  For $4/3<\gamma<3/2$, the number of nodes with three links grows as
$t^{3-2\gamma}$ and the number with more than three is finite.  Generally for
${m+1\over m}<\gamma<{m\over m-1}$, the number of nodes with more than $m$
links is finite, while $N_k\sim t^{k-(k-1)\gamma}$ for $k\leq m$.
Logarithmic corrections also arise at the transition points.

\subsection{Relation to citation data}

Let us now attempt to relate some of our predictions from the GN model to the
distribution of citations in recent scientific publications\cite{sor,redner}.
The GN model represents an extreme idealization of the citation process in
which each publication cites only a single paper and the probability of
citing a paper depends only on its current number of citations, and not on
its intrinsic quality or any other realistic features.  Thus we anticipate
that the connection between the model and the data will be, at best, tenuous.

The data that we discuss is based on: (a) 783,339 papers with 6,716,198
citations (provided by the Institute of Scientific Information (ISI)), and
(b) 24,296 papers with 351,872 citations from all issues of Physical Review D
(PRD) from 1975--1994 (provided by the SPIRES database)\cite{site}.  A
cursory visual inspection of this data suggests that the number of
publications with $k$ citations decays as a stretched exponential function of
$k$ (see {\it e.~g.}, Fig.~1 of Ref.~\cite{redner}).  However, an analysis
based on presenting the data in a Zipf plot, in conjunction with scaling, is
suggestive of a power-law form for the citation distribution, $k^{-\nu}$,
with $\nu\approx 3$ (Fig.~2 of Ref.~\cite{redner}).  This ambiguity between a
stretched exponential and power-law form for the citation distribution
corresponds to the situation where the predictions of the GN itself are
difficult to discern numerically.

If we consider the GN with attachment kernel $A_k\sim k^\gamma$ for
$\gamma\alt 1$, then a plot of $n_k$ in Eq.~(\ref{cases}) versus $k$, for
$1\leq k\leq 1000$, changes relatively slowly as $\gamma$ varies in the range
$(0.9,1)$.  If one attempts to fit this data to a power law, then an exponent
value somewhat larger than 3 gives a reasonable fit to the data.  It is only
as $\gamma\to 1$ from below, however, that the factors in the exponential of
Eq.~(\ref{cases}) conspire to give a pure power-law form for $n_k$.  Because
of the relatively small change in $n_k$ as $\gamma$ varies, the relatively
incomplete data on the distribution of citations is insufficient to provide a
clear test for the existence of a power law.  Further, for the GN model with
linear attachment kernel, the degree distribution depends on additional
details of this kernel and can achieve {\em any} value greater than 2.  In
short, it is difficult to relate the GN model to citation data based on the
form of the distribution alone.

Another interesting aspect of the citation distribution which can be compared
with the GN model is the nature of highly-cited publications.  Within the GN
model, the degree of the most popular node, $k_{\rm max}$, may be determined
by the extreme statistics criterion $\sum_{k>k_{\rm max}} N_k=1$, which
states that there is one node in the network whose degree lies in the range
$(k_{\rm max},\infty)$.  This criterion gives
\begin{equation}
\label{kmax}
k_{\rm max}\sim \cases{ 
   (\ln t)^{1/(1-\gamma)}  & $0\leq \gamma<1$; \cr
   t^{1/(\nu-1)}           & asymptotically linear; \cr  
   t                       & $\gamma>1$.}
\end{equation}
We now compare this prediction with the data about the most-cited paper.  To
make a correspondence between citations and Eq.~(\ref{kmax}), we identify the
total number of publications in each dataset with $t$.  The most cited paper
had 8,904 citations in the ISI data set and 2,026 citations in the PRD data
set.  These results are consistent with the first line of Eq.~(\ref{kmax})
when $\gamma\approx 0.86$ and $\gamma\approx 0.7$ respectively, and also with
the second line for $\nu\approx 2.5$ and $\nu\approx 2.3$ respectively.  Thus
an analysis of the most-cited paper does not cleanly indicate whether the
citation distribution is a power law or a stretched exponential.

These ambiguities indicate some of the issues that should be be clarified to
provide a clear description of citations in terms of a growing network model.

\subsection{GNR Model}

We now solve the GNR model within the rate equation framework.  According to
the basic processes in the model (Fig.~\ref{R}), the degree distribution
$N_k(t)$ evolves by the rate equations
\begin{eqnarray}
\label{NkC}
{d N_k\over dt}=&&\delta_{k1}
+{1-r\over M_0}\left[N_{k-1}-N_k\right]\nonumber\\
&&+{r\over M_0}\left[(k-2)N_{k-1}-(k-1)N_k\right].
\end{eqnarray}
For re-direction probability $r=0$, the first three terms on the right-hand
side of Eqs.~(\ref{NkC}) are the same as in the GN.  The last two terms
account for the change in $N_k$ due to the re-direction process.  To
understand their origin, consider the gain term due to re-direction.  Since
the initial node is chosen uniformly, if re-direction does occur, the
probability that a node with $k-1$ pre-existing links receives the new
``re-directed'' link is proportional to $k-2$, the number of pre-existing
incoming links.  A similar argument applies for the re-direction-driven loss
term.  Since $N_0\equiv 0$ is tacitly assumed, Eq.~(\ref{NkC}) applies for
all $k\geq 1$.

By combining the terms in Eq.~(\ref{NkC}), the rate equation reduces to that
of the original GN with $A_k=(k-1)r+1-r= r[k-1+(1-r)/r]$.  By scaling out the
factor $r$, we then reduce $A_k$ to the shifted linear kernel $k+w$, with
$w=(1-r)/r-1={1\over r}-2$.  Thus we can merely transcribe our results about
the GN with the shifted linear kernel to determine the degree distribution
for the GNR model.  Amusingly, for $r=1/2$, the $GNR$ model is identical to
the GN with the purely linear kernel.  In general, the degree distribution in
the $R$ model is a power law with exponent $\nu=1+1/r$, which can be tuned to
any value larger than 2.  This exponent value was first obtained in Simon's
original paper\cite{simon}, but in a rather different context, by employing
an approach which is similar to ours.

\section{The age distribution}

In addition to the distribution of degree, we study {\em when}
connections occur in the GN.  This provides a deeper understanding of
the overall development of growing networks.  Naively, we expect that
older nodes will be better connected and this can be quantified by
categorizing nodes both by their degree and their age.  It should be
emphasized, that the GN does {\em not\/} have explicit aging, in which
the connection probability depends on the age of the target node; this
feature is treated in Ref.\cite{j1}.  Instead, we are merely extending
the categorization of node to include their age as well as their degree.

\subsection{Linear connection kernel}

Let $c_k(t,a)$ be the average number of nodes of age $a$ which have $k-1$
incoming links at time $t$.  Here age $a$ means that the node was introduced
at time $t-a$.  That is, we are now resolving each node both by its degree
and its age.  The resulting joint age-degree distribution is simply related
to the degree distribution through $N_k(t)=\int_0^t da\,c_k(t,a)$.  The joint
distribution evolves according to
\begin{equation}
\label{ck}
\left({\partial \over \partial t}+{\partial \over \partial a}\right)c_k 
={A_{k-1}c_{k-1}-A_k c_k\over A(t)}
+\delta_{k1}\delta(a).
\end{equation}
The second term on the left accounts for the aging of nodes and the
probability of connecting to a given node again depends only on its degree
and not on its age.

We start by considering the linear attachment kernel, $A_k=k$, and focus
on the long time asymptotic behavior.  Then we can disregard the initial
condition and write $A(t)\equiv M_1(t)=2t$.  This transforms
Eqs.~(\ref{ck}) into
\begin{equation}
\label{ck1}
\left({\partial \over \partial t}+{\partial \over \partial a}\right)c_k 
={(k-1) c_{k-1}-k c_k\over 2t}+\delta_{k1}\delta(a).
\end{equation}
The homogeneous form of this equation implies that solution should be
self-similar.  Thus we seek a solution as a function of the {\em single}
variable $a/t$ rather than two separate variables.  Thus, we write
\begin{equation}
\label{ck1scal}
c_k(t,a)=f_k(x) \qquad {\rm with}\quad x=1-{a\over t}.
\end{equation}
This turns the partial differential equation (\ref{ck1}) into the
ordinary differential equation
\begin{equation}
\label{fk1}
-2x\,{df_k\over dx}=(k-1) f_{k-1}-k f_k.
\end{equation}
We have omitted the delta function term, since it merely provides the
boundary condition $c_k(t,a=0)=\delta_{k1}$, or
\begin{equation}
\label{fkbound}
f_k(1)=\delta_{k1}.
\end{equation}

\begin{figure}
  \narrowtext \hskip 0.1in \epsfxsize=2.8in
\epsfbox{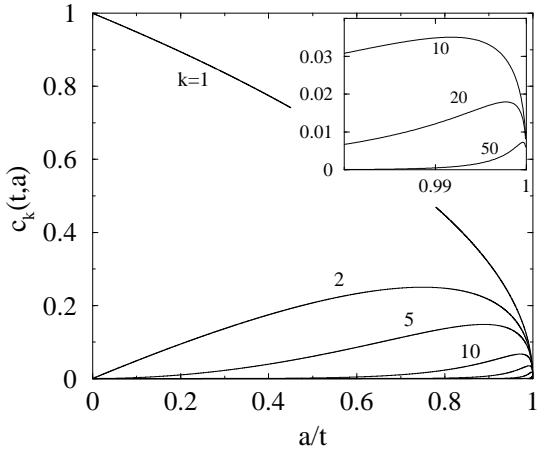} \vskip 0.1in
\caption{Age-dependent degree distribution for the GN for the linear 
  attachment kernel.  Low-degree nodes tend to be relatively young while
  high-degree nodes are old.  The inset shows detail for $a/t\geq 0.98$.
\label{age}}
\end{figure}

The solution to this boundary-value problem may be simplified by assuming the
exponential solution $f_k=\Phi\varphi^{k-1}$; this is consistent with the
boundary condition, provided that $\Phi(1)=1$ and $\varphi(1)=0$.  The above
ansatz reduces the infinite set of rate equations (\ref{fk1}) into two
elementary differential equations for $\varphi(x)$ and $\Phi(x)$ whose
solutions are $\varphi(x)=1-\sqrt{x}$ and $\Phi(x)=\sqrt{x}$.  In terms of
the original variables of $a$ and $t$, the joint age-degree distribution is
then
\begin{eqnarray}
\label{ck1all}
c_k(t,a)=\sqrt{1-{a\over t}}\left\{1-\sqrt{1-{a\over t}}\right\}^{k-1}.
\end{eqnarray}

Thus the degree distribution for nodes of fixed age decays exponentially with
degree, with a characteristic degree which diverges as $\langle k\rangle\sim
(1-a/t)^{-1/2}$ for $a\to t$.  As expected, young nodes (those with $a/t\to
0$) typically have a small degree while old nodes have large degree
(Fig.~\ref{age}).  It is the slow decay of the degree distribution for old
nodes which ultimately leads to a power-law degree distribution when this
joint age-degree distribution is integrated over all ages to give $N_k(t)$.

\subsection{General connection kernels}

Let us now consider the GN with a connection kernel which grows either
linearly or more slowly with $k$.  The ansatz (\ref{ck1scal}) still is valid,
so that the distribution $f_k$ evolves according to
\begin{equation}
\label{fk}
-\mu x\,{df_k\over dx}
=A_{k-1} f_{k-1}-A_k f_k.
\end{equation}
We now solve Eqs.~(\ref{fk}), subject to the boundary condition
(\ref{fkbound}), and with $\mu$ determined from Eq.~(\ref{mugen}).  Let us
first replace $x$ by $X=-\mu^{-1}\ln x$, which reduces the left-hand side of
(\ref{fk}) to ${df_k\over dX}$.  Applying a Laplace transform, $\hat
f_k(s)=\int_0^\infty dX\,e^{-sX}f_k(X)$, $\hat f_k(s)$ obeys a simple
algebraic recursion formula whose solution is
\begin{eqnarray}
\label{fk-sol}
\hat f_k(s)={1\over A_k}\prod_{j=1}^k \left(1+{s\over A_j}\right)^{-1}.
\end{eqnarray}

Apart from notation, this is identical to Eq.~(\ref{Nkgen}) and can be
analyzed accordingly.  In particular, we can determine $\hat f_k(s)$ for
various asymptotically linear attachment kernels.  For example, for the
shifted linear attachment kernel, $A_k=k+w$, we find
\begin{equation}
\label{fkw}
\hat f_k(s)={\Gamma(1+w+s)\over \Gamma(1+w)}\,
{\Gamma(k+w)\over \Gamma(k+1+w+s)}.
\end{equation}
To invert this Laplace transform, it is useful to rewrite this expression as
a sum of rational functions $\hat f_k(s)=\sum_{1\leq j\leq k} F_j^k
(j+w+s)^{-1}$.  This then gives $f_k(X)=\sum_{1\leq j\leq k}
F_j^k\,e^{-(j+w)X}$, with
\begin{equation}
\label{Fjk}
F_j^k={(-1)^{j-1}\Gamma(k+w)\over \Gamma(j)\,\Gamma(k-j+1)\,\Gamma(1+w)}.
\end{equation}
When then re-express this in terms of the original variable $x=e^{-(2+w)X}$.
Hence $f_k(x)$ can be re-written as the sum of $k$ power-laws
$f_k(x)=\sum_{1\leq j\leq k} F_j^k\, x^{(j+w)/(2+w)}$.  Substituting the
explicit expressions (\ref{Fjk}) into this sum reduces the joint
age-degree distribution to
\begin{equation}
\label{fkwx}
f_k(x)={\Gamma(k+w)\over \Gamma(k)\,\Gamma(1+w)}\,\,x^{1+w\over 2+w}
\left[1-x^{1\over 2+w}\right]^{k-1}.
\end{equation}

This expression shows that old nodes have a broad distribution of degrees up
to a characteristic degree $\langle k\rangle= (1-a/t)^{-1/(2+w)}$.  One can
also verify that the average age $a_k$ of nodes of degree $k$, defined as
$a_k=N_k^{-1}\int_0^t da\,a c_k(t,a)=tn_k^{-1}\int_0^1 dx\,(1-x)f_k(x)$, is
\begin{eqnarray}
\label{akav}
{a_k\over t} &=&1-{\Gamma(5+3w)\,\Gamma(k+3+2w)\over 
\Gamma(3+2w)\,\Gamma(k+5+3w)}\nonumber \\
  &\sim& 1-{{\rm const.}\over k^{2+w}}.
\end{eqnarray}
Thus nodes with very large degree necessarily have an age which approaches
that of the entire network.

Finally, the joint age-degree distribution simplifies in the limit
$k\to\infty$ and $x\to 0$, with the scaling variable $\xi=kx^{1/(2+w)}$ kept
finite.  In this case, we can rewrite (\ref{fkwx}) in the scaling form
\begin{equation}
\label{ckw-sol}
f_k(x)=k^{-1}F(\xi), \quad 
F(\xi)={\xi^{1+w}\over \Gamma(1+w)}\,\,\exp(-\xi).
\end{equation}
The scaling variable can also be written as $\xi=k/\langle k\rangle$, and
thus Eq.~(\ref{ckw-sol}) clearly shows that old nodes have a broad
distribution of degrees: $1\leq k\alt \langle k\rangle$.

We can derive explicit age-degree distributions for other attachment
kernels.  For example, for the constant attachment kernel, $A_k=1$, the joint
age-degree distribution is the Poisson distribution,
\begin{equation}
\label{fk0}
f_k(X)={X^{k-1}\over (k-1)!}\,e^{-X},
\end{equation}
or in terms of the original variables $a$ and $t$,
\begin{equation}
\label{ck0}
c_k(t,a)=\left(1-{a\over t}\right)
{|\ln(1-a/t)|^{k-1}\over (k-1)!}.
\end{equation} 
The characteristic degree now diverges relatively slowly, {\it viz.} $\langle
k\rangle\sim -\ln(1-a/t)$ as $a\to t$, than for asymptotically linear
attachment kernels.  On the other hand, the average age approaches the
maximal age $t$ at a much faster rate, as $a_k=t\left[1-(2/3)^k\right]$, as
$k$ approaches its maximal value.

For cases where we have been unable to obtain an explicit solution, the
Laplace transform method still allows us to extract the asymptotics.  For
example, for asymptotically homogeneous attachment kernels, $A_k\to
k^{\gamma}$ as $k\to\infty$, Eq.~(\ref{fk-sol}) gives the large-$k$
asymptotics $\hat f_k(s)\sim k^{-\gamma}
\exp\left[-sk^{1-\gamma}/(1-\gamma)\right]$ (see Eq.~(\ref{cases}).  (For
concreteness, we consider here the range $1/2<\gamma<1$.)~ Inverting this
Laplace transform yields
\begin{equation}
\label{fkg}
f_k(X)\sim k^{-\gamma}
\delta\left(X-{k^{1-\gamma}\over 1-\gamma}\right).
\end{equation}
In particular, the age of nodes with $k$ links is peaked about the value
$a_k$ which satisfies
\begin{equation}
\label{akg}
{a_k\over t} \simeq 
1-\exp\left(-\mu\,{k^{1-\gamma}\over 1-\gamma}\right).
\end{equation}
This again shows that old nodes are much better connected.

\section{Node Degree Correlations}

We now demonstrate that correlations between the degrees of connected nodes
spontaneously develop as the network grows.  One motivation for focusing on
these correlations is that recently random graph models with arbitrary degree
distributions have been investigated\cite{reed,chung,newman}.  While the
degree distribution can be chosen arbitrarily in these models, the degrees of
connected nodes are {\em uncorrelated}.  This lack of correlation suggests
that such random graphs may have limited applicability to growing network
systems.

For the GN, a useful characterization of node degree correlations is
$N_{kl}(t)$, the number of nodes of total degree $k$ which attach to an
ancestor node of total degree $l$.  For example, in the network of
Fig.~\ref{network}, there are $N_1=6$ nodes of degree 1, with
$N_{12}=N_{13}=N_{15}=2$.  There are also $N_2=2$ nodes of degree 2, with
$N_{25}=2$, and $N_3=1$ node of degree 3, with $N_{35}=1$.  The correlation
function is not defined for the initial node.  Generally, $N_{kl}$ is defined
for $k\geq 1$ and $l\geq 2$, and obeys the sum rule $N_k=\sum_l N_{kl}$.  A
gratifying feature of the rate equation approach is that the correlation
function $N_{kl}$ can be understood in a natural and simple fashion.

\subsection{Linear connection kernel}

For the GN with the linear attachment kernel $A_k=k$, the joint distribution
$N_{kl}(t)$ evolves according to
\begin{eqnarray}
\label{Nkl}
M_1\,{d N_{kl}\over dt}&=&
\left[(k-1) N_{k-1,l}-kN_{kl}\right]+\nonumber \\
&& \left[(l-1) N_{k,l-1}-l N_{kl}\right]+(l-1)N_{l-1}\,\delta_{k1}.
\end{eqnarray}
The first two terms on the right-hand side account for the change in $N_{kl}$
due to the addition of a link onto a node of degree $k-1$ (gain) or $k$
(loss), while the second set of terms gives the change in $N_{kl}$ due
to the addition of a link onto the ancestor node.  Finally, the last term
accounts for the gain in $N_{1l}$ due to the addition on the new node.  

Asymptotically, $M_1\to 2t$ and $N_{kl}\to tn_{kl}$, and we use these
hypotheses to reduce Eqs.~(\ref{Nkl}) to the time-independent recursion
relations
\begin{eqnarray}
\label{nkl}
(k+l+2)n_{kl}&=&(k-1) n_{k-1,l}+(l-1) n_{k,l-1}\nonumber\\
&+&(l-1)n_{l-1}\,\delta_{k1}.
\end{eqnarray}
This can be reduced to a constant-coefficient inhomogeneous recursion
relation by the substitution
\begin{equation}
\label{Akl}
n_{kl}={\Gamma(k)\,\Gamma(l)\over \Gamma(k+l+3)}\,\,m_{kl}
\end{equation}
to yield
\begin{equation}
\label{A}
m_{kl}=m_{k-1,l}+m_{k,l-1}+4(l+2)\delta_{k1}.
\end{equation}
By solving Eqs.~(\ref{A}) for the first few $k$, one can grasp the pattern of
dependence on $k$ and $l$ and thereby infer the general solution
\begin{equation}
\label{A-sol}
m_{kl}=4\,{\Gamma(k+l)\over \Gamma(k+2)\,\Gamma(l-1)}
+12\,{\Gamma(k+l-1)\over \Gamma(k+1)\,\Gamma(l-1)}.
\end{equation}
This solution can also be obtained in a more systematic manner by the
generating function method (see below for the shifted linear kernel).
Combining Eqs.~(\ref{Akl}) and (\ref{A-sol}) we finally obtain
\begin{eqnarray}
\label{nkl-sol}
n_{kl}&=&{4(l-1)\over k(k+1)(k+l)(k+l+1)(k+l+2)}\nonumber\\
&+&{12(l-1)\over k(k+l-1)(k+l)(k+l+1)(k+l+2)}.
\end{eqnarray}

The important feature of this result is that the joint distribution does not
factorize, that is, $n_{kl}\ne n_kn_{l}$.  This confirms our earlier
assertion that correlations between the degrees of connected nodes form
spontaneously.  This is arguably the most important distinction between
classical random graphs -- where node degrees are uncorrelated -- and the
GN.

While the solution of Eq.~(\ref{nkl-sol}) is unwieldy, it greatly simplifies
in the scaling regime, $k\to\infty$ and $l\to\infty$ with $y=l/k$ kept
finite.  The scaled form of the solution is
\begin{eqnarray}
\label{nkl-scal}
n_{kl}=k^{-4}\,{4y(y+4)\over (1+y)^4}.
\end{eqnarray}
For fixed large $k$, the distribution $n_{kl}$ has a single maximum at
$y_*=(\sqrt{33}-5)/2 \cong 0.372$.  Thus a node whose degree $k$ is large is
typically linked to another node whose degree is also large; the typical
degree of the ancestor is 37\% of the degree of the daughter node.  In the
complementary case of a fixed degree $l$ for the ancestor node, the
distribution $n_{kl}$ reaches maximum when $k=1$, {\it i.\ e.}, the daughter
node is usually dangling.  {}From Eq.~(\ref{nkl-sol}), we find that this
configuration occurs with probability
\begin{eqnarray}
\label{n1l}
n_{1l}={2(l-1)(l+6)\over l(l+1)(l+2)(l+3)}.
\end{eqnarray}
Finally, when both $k$ and $l$ are large and also their ratio is very
different from one, the limiting behaviors of $n_{kl}$ are
\begin{equation}
\label{nklext}
n_{kl}\to\cases{16\,(l/k^5)    & when $l\ll k$,\cr
                4/(k^2\,l^2)   & when $l\gg k$.\cr}
\end{equation}
This last result demonstrates the correlations in the network most cleanly.
If there were no correlations, then $n_kn_l$ would be proportional to
$(k\,l)^{-3}$.

\subsection{General connection kernels}

In general, correlations between the degrees of neighboring connected nodes
exist for any attachment kernel.  The analysis of these correlations for an
arbitrary kernel is tedious and we merely outline some of the primary results
in the relatively simple cases of the shifted linear and constant attachment
kernels.  

In the former case, we follow the same approach as the linear kernel to
reduce the rate equation for the correlation function to recursion relations
of a similar form to Eq.~(\ref{nkl}), {\it viz.}
\begin{eqnarray}
\label{nklw}
(k+l+2+3w)&&n_{kl}=(k+w-1) n_{k-1,l}+\\
&&(l+w-1)\left[n_{k,l-1}+n_{l-1}\,\delta_{k1}\right].\nonumber
\end{eqnarray}
Here $n_l$ is determined from Eq.~(\ref{nkw}).  In analogy with
Eq.~(\ref{Akl}), the substitution
\begin{equation}
\label{mKL}
n_{kl}={\Gamma(k+w)\,\Gamma(l+w)\over \Gamma(k+l+3+3w)}\,\,m_{kl}
\end{equation}
reduces Eqs.~(\ref{nklw}) to
\begin{equation}
\label{mW}
m_{kl}=m_{k-1,l}+m_{k,l-1}+\delta_{k1}\,W\,\,
{\Gamma(l+3+3w)\over
\Gamma(l+2+2w)},
\end{equation}
where $W=(2+w)\,\Gamma(3+2w)/(\Gamma(1+w))^2$.  We solve the recursion
(\ref{mW}) by the generating function method\cite{knuth}.  Multiplying
Eq.~(\ref{mW}) by $x^ky^l$ and summing over all $k\geq 1, l\geq 2$, we find
that the generating function
\begin{equation}
\label{Mxy}
{\mathcal M}(x,y)=\sum_{k=1}^\infty \sum_{l=2}^\infty m_{kl}x^ky^l 
\end{equation}
is given by
\begin{equation}
\label{Mxy-sol}
{\mathcal M}(x,y)={Wxy^2\over 1-x-y}\sum_{j=0}^\infty 
{\Gamma(j+5+3w)
\over \Gamma(j+4+2w)}\,\,y^j.
\end{equation}
Expanding ${\mathcal M}(x,y)$ we obtain
\begin{equation}
\label{mKL-sol}
m_{kl}=W\sum_{j=0}^{l-2} 
{\Gamma(k+l-2-j)\,\Gamma(j+5+3w)\over 
\Gamma(k)\,\Gamma(l-1-j)\,\Gamma(j+4+2w)}.
\end{equation}
Eqs.~(\ref{mKL}) and (\ref{mKL-sol}) constitute the exact solution for the
correlation function of the GN with the shifted linear attachment kernel.

When the parameter $w$ is an integer, we can reduce $n_{kl}$ to a rational
function.  In the general case, the exact solution also simplifies in several
extreme limits.  When $k\gg l$, the dominant contribution to $n_{kl}$ is
provided by the first term in the sum in Eq.~(\ref{mKL-sol}).  Assuming
additionally $l\gg 1$ and repeatedly using the asymptotic relation
$\Gamma(N+n)/\Gamma(N)\to N^n$ as $N\to\infty$, we ultimately find
\begin{equation}
\label{nklw1}
n_{kl}\simeq W\,{\Gamma(5+3w)\over \Gamma(4+2w)}\,\,l^{1+w}\,k^{-5-2w}\qquad
k\gg l\gg 1.
\end{equation}
In the complementary case of $l\gg k\gg 1$, all the terms in the sum of
Eq.~(\ref{mKL-sol}) are important.  However, we can simplify this sum by
employing the above asymptotics for the ratio of gamma functions and then
replacing the sum by an easily-computable integral.  We find
\begin{equation}
\label{nklw2}
n_{kl}\simeq W\,\Gamma(2+w)\,k^{-2}\,l^{-2-w}.
\end{equation}

When the attachment kernel is uniform, correlations between the degrees of a
node and its ancestor still develop.  To see how this comes about
quantitatively, we again follow the same steps as those which led to
Eq.~(\ref{nkl}) and find that the joint distribution $n_{kl}$ now satisfies
the recursion relation
\begin{eqnarray}
\label{nkl0}
3n_{kl}=n_{k-1,l}+n_{k,l-1}+2^{-(l-1)}\delta_{k1}.
\end{eqnarray}
This recursion relation can again be solved by the generating function
technique to give
\begin{eqnarray}
\label{nkl0-sol}
n_{kl}={1\over 2^{l-1}}-{1\over 3^{l-1}}\sum_{i=0}^{k-1}
{\Gamma(l-1+i)\over \Gamma(l-1)\,\Gamma(i+1)}\,{1\over 3^i}.
\end{eqnarray}
To appreciate the qualitative behavior of the joint distribution $n_{kl}$ it
is again useful to fix one variable and vary the other.  For fixed $l$,
Eq.~(\ref{nkl0-sol}) shows that $n_{kl}$ has a maximum at $k=1$.  The
magnitude of this maximum is $n_{1l}=2^{-(l-1)}-3^{-(l-1)}$.  To analyze the
behavior when $k$ is fixed, it is convenient to transform
Eq.~(\ref{nkl0-sol}) into
\begin{equation}
\label{nkl0-sol2}
n_{kl}={1\over 3^{l-1}}\sum_{i=k}^\infty
{\Gamma(l-1+i)\over \Gamma(l-1)\,\Gamma(i+1)}\,{1\over 3^i}.
\end{equation}
Now a straightforward analysis shows that for large $k$, the maximum is
attained at $l=k/2$.

The form of the joint distribution $n_{kl}$ remains relatively complex even
in the scaling regime, $k,l\to\infty$, with the scaling variable $y=l/k$ kept
finite.  We determine the scaled form of the solution (\ref{nkl0-sol2}) by
applying Stirling's formula and the identity $\Gamma(x+\lambda)/\Gamma(x)\to
x^\lambda$ as $x\to\infty$.  For $y<2$, we find
\begin{equation}
\label{nkl0-scal}
n_{kl}\simeq {1\over \sqrt{2\pi k}}\,{\sqrt{1+y^{-1}}\over 2-y}\,e^{-kY},
\end{equation}
where $Y=y\ln y-(y+1)\ln[(y+1)/3]$.  For $y>2$, it is preferable to use
the solution in the form of Eq.~(\ref{nkl0-sol}).  After some algebra,
we can verify that the dominant contribution equals $2^{-(l-1)}$,
that is, {\em independent} of $k$\cite{note1}.  

Finally, the limiting behavior of the correlation function is
\begin{equation}
\label{next}
n_{kl}\to 2^{-1}\times \cases{
3^{-(k+l-2)}\,{k^{l-2}\over (l-2)!}   & when $l\ll k$,\cr
2^{-(l-2)}                            & when $l\gg k$.\cr}
\end{equation}
Thus, correlations are strong even for the random attachment kernel and the
qualitative behavior is similar to that of the linear attachment kernel.

\section{Large-Scale Properties}

\begin{figure}
  \narrowtext \hskip 0.1in \epsfxsize=3.in
\epsfbox{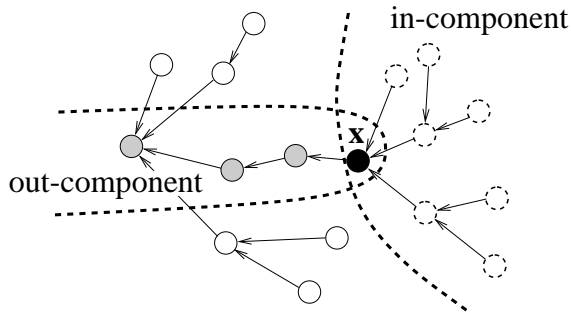} \vskip 0.1in
\caption{In-component and out-component of node {\bf x}.
\label{in-out}}
\end{figure}

The degree of a node is an important but local network characteristic and we
now seek to quantify more global features of the network.  One such
characteristic is the partitioning of the network into an {\em
  in-component\/} and an {\em out-component\/} with respect to any node
(Fig.~\ref{in-out}).

The in-component to node {\bf x} is the set of all nodes from which node {\bf
  x} can be reached by following a path of directed links.  Similarly the
out-component of node {\bf x} is the set of nodes which can be reached by
following the path directed links which emanate from node {\bf x}.  For the
GN model, the out-component is just a single path, while in more realistic
networks both the in- and out-components will be branched.  In the context of
citations, the in-component is the set of all publications which refer to
{\bf x}, either directly or through intermediate reference lists until {\bf
  x} is reached.  The out-component is the set of cited publications
generated by iteratively following the reference list(s) of {\bf x} and its
ancestors.

\subsection{In-component size distribution}

The size distribution of the in-component can be easily obtained by the
rate equation formalism for the GN with a uniform attachment kernel and
also for the GNR.  Given the equivalence between the latter and the GN
with a shifted linear kernel, the latter case is also soluble.  We start
by considering the GN with the uniform attachment kernel.  In this case,
the number $I_s(t)$ of in-components with $s$ nodes satisfies the rate
equation
\begin{equation}
\label{Ik}
{d I_s\over dt}={(s-1)I_{s-1}-sI_s\over t}+\delta_{s1}.
\end{equation}
To understand this equation, consider first the loss term.  For an
in-component of size $s$ there are $s$ nodes in which the attachment of a new
node causes this component to increase in size by one.  This gives a loss
rate for $I_s$ which is proportional to $s$.  If there is more than one
in-component of size $s$ they must be disjoint, so that the total loss rate
for $I_s$ is simply $sI_s$.  A similar argument applies for the gain term.
Finally, the overall factor of $t^{-1}$ converts these rates to normalized
probabilities.  Curiously, Eqs.~(\ref{Ik}) are almost identical to the rate
equations for the degree distribution of the GN with linear attachment
kernel, except that the prefactor equals $t^{-1}$ rather than $(2t)^{-1}$.

{}From Eqs.~(\ref{Ik}) we can determine all moments of the in-component size
distribution, ${\mathcal I}_n(t)=\sum_{s\geq 1} s^n I_s(t)$.  The zeroth
moment obeys $\dot {\mathcal I}_0=1$, whose solution is ${\cal
  I}_0(t)={\mathcal I}_0(0)+t$.  This is obvious since the total number of
in-components equals the total number of nodes.  The first moment obeys $\dot
{\cal I}_1=1+{\cal I}_1/{\cal I}_0$, whose asymptotic solution is ${\mathcal
  I}_1(t)\sim t\ln t$.  We shall see that this logarithmic factor is an
outcome of the asymptotic power law for $I_s$ with the tail decaying as
$s^{-2}$.

To solve for $I_s(t)$, we note that it again grows linearly in time.  Thus we
substitute the ansatz $I_s(t)=ti_s$ into Eqs.~(\ref{Ik}) to obtain $i_1=1/2$
and $i_s=i_{s-1}(s-1)/(s+1)$, which immediately leads to
\begin{equation}
\label{is}
i_s={1\over s(s+1)}.  
\end{equation}
For this $s^{-2}$ decay, the moments ${\cal I}_n$ diverge when $n\geq 1$.
However, the size of the largest in-component, $s_{\rm max}=t$, provides an
upper threshold in the computation of the moments.  For example, ${\cal
  I}_1\sim \sum_{s\leq t} s I_s(t)=t\ln t$.  It is intriguing that the
algebraic in-component distribution co-exists with an exponential in-degree
distribution, $n_k=2^{-k}$.

Similarly, we can determine $I_s(t)$ for the GNR model.  In this case, the
number $I_s(t)$ of in-components with $s$ nodes satisfies
\begin{equation}
\label{IkC}
{d I_s\over dt}={(s-2+(1-r))I_{s-1}-(s-1+(1-r))I_s\over t}
\end{equation}
for $s\geq 2$, and $\dot I_1=1-(1-r)I_1/t$.  This rate equation can be
understood in a similar manner as Eq.~(\ref{Ik}).  Consider the loss term for
an in-component of size $s$.  There are two possibilities to consider: (i) If
the apex of the in-component is initially chosen, then the new node will
attach to this apex with probability $1-r$ ({\it i.\ e.}, attach with no
re-direction); (ii) If any other of the $s-1$ nodes of the in-component is
chosen, the new node will surely attach to the in-component even if
re-direction occurs.  These two processes give a loss rate for $I_s$ which is
proportional to $(s-1+(1-r))I_s$.  Solving for the in-component distribution
in this process now yields $I_s(t)=ti_s$, with
\begin{equation}
\label{qkC}
i_s={1-r\over (s-r)(s+1-r)}.
\end{equation}
Remarkably, the asymptotic power law $I_s\propto s^{-2}$ holds for any $r$.
It is striking that this apparently universal behavior has also recently
been observed in measurements of the Internet\cite{cald}

Since the GNR model is identical to the GN with the shifted linear attachment
kernel $A_k=k+({1\over r}-2)$, Eq.~(\ref{qkC}) also applies to the
in-component distribution for the GN with shifted linear attachment kernels.
For example, the in-component distribution for the linear kernel is
$i_s=2/(4s^2-1)$.  Since the same $I_s\propto s^{-2}$ decay holds for the GN
with both constant and linear attachment kernels, we conjecture that the
in-component distribution exhibits a universal $s^{-2}$ decay for an {\em
  arbitrary} attachment kernel, as long as it does not grow faster than
linearly with node degree.

\subsection{Out-component size distribution}

The out-component from each node reveals basic insights about the
``genealogy'' of the growing network in an extremely simple fashion.  For
example, it allows us to estimate the diameter of the network, an important
characteristic which has been measured for the web graph\cite{b3,brod} and
for social networks\cite{small}.

\begin{figure}
  \narrowtext \hskip 0.2in \epsfxsize=2.6in
\epsfbox{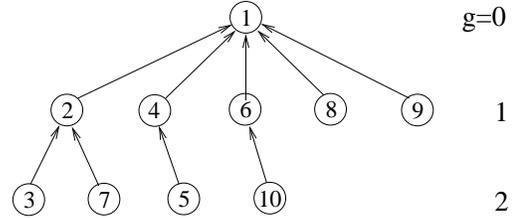} \vskip 0.1in
\caption{Genealogy of the growing random network of 
  Fig.~\ref{network}.  The indices indicate when a node is introduced, while
  the ancestor determines where a new node is positioned.
\label{genealogy}}
\end{figure}

For this characterization, we begin by reorganizing the GN into a
genealogical tree according to a procedure which is suggested by the growth
process itself.  Generation $g=0$ contains the single ``seed'' node.  The
nodes which attach to the seed node form generation $g=1$, and generally the
nodes which attach to nodes in generation $g$ form generation $g+1$, {\em
  independent\/} of when the attachment actually occurs.  Thus the position
of a node in the genealogical tree depends only on the position of the
ancestor node and {\em not\/} on when the node is introduced.  In this
respect, the GN genealogical tree differs from usual genealogies, where each
new generation is born into a progressively later position in the
genealogical tree.  For example, the network of Fig.~1 has 5 nodes in the
first generation and 4 nodes in the second generation leading to the
genealogical tree of Fig.~\ref{genealogy}.  The sizes of all generations grow
continuously, except for generation $g=0$ which always consists of the single
node.

Once we understand the genealogical structure of the GN, we
simultaneously establish the out-component distribution.  Indeed, the
number $O_s$ of out-components with $s$ nodes equals $L_{s-1}$, the
number of nodes in generation $s-1$ in the genealogical tree.  We
therefore compute $L_g(t)$, the size of generation $g$ at time $t$.  We
start with the simplest situation when the attachment rate is uniform.
In this case, $L_g(t)$ increases when a new node attaches to a node in
the previous generation.  This occurs with rate $L_{g-1}/M_0$, where
$M_0(t)=1+t$ is the total number of nodes.  Because of the simplicity of
the corresponding rate equations, we use the exact expression for $M_0$
rather than the asymptotic expression $M_0\sim t$, as was done in
solving for the in-component.  Thus we write
\begin{equation}
\label{L}
{d L_g\over dt}={L_{g-1}\over 1+t}.
\end{equation}
Solving these equations gives
\begin{equation}
\label{Lsol}
L_g(\tau)={\tau^g\over g!}\qquad {\rm where} \quad \tau=\ln(1+t).
\end{equation}
We therefore conclude that for a fixed (large) time, the generation size
grows with $g$ when $g<\tau$, reaches a maximum size which is equal to
\begin{equation}
\label{Lmax}
L_{\rm max}\simeq {t\over \sqrt{2\pi \ln t}}
\end{equation}
when $g=\tau$, and then decreases and eventually becomes of order one when
$g=e\tau$.  The distribution $L_g$ quickly decays when $g$ exceeds the cutoff
value $e\tau$.  At time $t$, the genealogical tree therefore contains
approximately $e\tau$ generations.  Hence the diameter $D$ of the network is
approximately $2e\tau$, or
\begin{equation}
\label{D}
D\approx 2e\ln N
\end{equation}
where $N=1+t$ is the total number of nodes.  Thus, the diameter of an {\em
  evolving} GN exhibits the same $N$ dependence as a {\em static} random
graph\cite{bol}.

We can also find the generation size distribution for shifted linear
attachment kernels.  It is again simpler to derive the rate equations in the
framework of the GNR model and then transcribe the results to the shifted
linear kernel.  For the GNR model, the rate equation for the generation size
distribution is
\begin{equation}
\label{LC}
{d L_{g}\over dt}={(1-r) L_{g-1}+r L_{g}\over 1+t}
\end{equation}
for $g> 1$, and $\dot L_1=(1+t)^{-1}[1+rL_1]$.  The first term in (\ref{LC})
has the same origin as in the GN without re-direction and the second term
accounts for the change in $L_g$ due to the re-direction.  In the latter
case, the new node provisionally attaches to a node in generation $g$; this
occurs with relative probability $L_g$.  However, by the re-direction
process, this new node actually attaches to a node in generation $g-1$ and
thereby joins generation $g$.

To solve Eq.~(\ref{LC}), we again use $\tau=\ln(1+t)$ and apply the Laplace
transform technique.  After some elementary steps, we obtain
\begin{equation}
\label{LCsol}
L_{g+1}(\tau)=\int_0^\tau dx\,{[(1-r) x]^g\over g!}\,e^{xr}.
\end{equation}
{}From this solution, we find that for a fixed (large) time, the generation
size grows with $g$ when $g<(1-r)\tau$, reaches a maximum value $L_{\rm
  max}\simeq t/\sqrt{2\pi (1-r)\ln t}$ at $g=(1-r)\tau$, and then decreases
when $g>(1-r) \tau$.  Eventually the generation size becomes of order one
when $g=G\tau$, where $G$ is the root of equation $G\ln(G/(1-r))=G+r$.  The
diameter of the network is then $D\approx 2G\tau$.

These two solvable cases again suggest that the genealogy of the GN is
robust, as long as the attachment kernel does not grow faster than linearly
with node degree.  For the super-linear kernels, however, the genealogy
changes drastically.  When the attachment exponent exceeds 2, there will be
only a few generations overall, and one generation $g^*$ will contain all but
the finite number of nodes.  For such a network, the gel node will reside in
generation $g^*-1$.  When the attachment exponent lies in the range
$1<\gamma<2$, a single generation will also contain almost all $t$ nodes.
However, the number of nodes which reside in other generations is of order
$t^{2-\gamma}$ and thus grows as well.  Additionally, the number of non-empty
generations grows indefinitely with the total number of nodes.

The above results can be reformulated in terms of the out-component
distribution.  In particular, for the GN with uniform attachment kernel, the
number $O_s$ of out-components with $s$ nodes equals
\begin{equation}
\label{Rk}
O_s(\tau)={\tau^{s-1}\over (s-1)!}\qquad {\rm where} \quad \tau=\ln(1+t).
\end{equation}
Similar results apply for the linear attachment kernel, suggesting that the
out-component distribution is robust as long as the attachment kernel does
not grow faster than linearly with node degree.

\section{Discussion and conclusions}

In this paper, we have analyzed the structure of the growing network (GN)
model and shown that many of its properties can be easily determined within a
rate equation approach.  We have found that the GN has a power-law node
degree distribution, $N_k(t)\sim tk^{-\nu}$, for asymptotically linear
attachment kernels, with an exponent $\nu$ which is always larger than 2.  By
tuning parameters of the model in a reasonable way, it is easy to obtain a
node degree distribution which is in quantitative agreement with available
data for the web graph\cite{kum1,klein,kum2,fal,b1,matta,brod}.

A remarkable feature of this network is the spontaneous development of
correlations between connected nodes.  These correlations provide a much more
sensitive characterization of the structure of growing networks than the
extensively studied degree distribution.  These correlations are a crucial
feature which distinguishes the GN from classical random graphs.  Thus
testing for the presence of correlations between node degrees in large
evolving networks may provide crucial insights to help determine the
underlying mechanism of their growth.

We have also studied two specific large-scale properties of the network,
namely, the size distributions of the in- and out-components with respect to
a given site.  The in-component distribution exhibits a robust $s^{-2}$
power-law behavior, where $s$ is the component size, as long as the
attachment probability does not grow faster than linearly with node degree.
The out-component distribution reveals the basic genealogical feature that
the number of ``generations'' in the network grows logarithmically with the
total number of nodes, again for attachment kernels which do not grow faster
than linearly in node degree.

The qualitative agreement between the degree distributions of real evolving
networks, such as the web graph, and the GN is reassuring given that the
model ignores many important features of real networks.  Nevertheless, a
number of characteristics of real growing networks are difficult to treat in
the framework of the GN model.  One important such characteristic is the
out-degree distribution.  Within the GN model the out-degree of each node is
one by construction.  In contrast, for real growing networks the out-degree
distribution has a power law form\cite{brod}.  Additionally, the average in-
and out-degrees at each node are generally larger than one.  For the web
graph, for example, $\langle i\rangle=\langle j\rangle\approx
7.5$\cite{brod}.

There are several natural ways to extend the GN model to generate an average
out-degree which is greater than one.  A simple construction is to link every
new node to more than one earlier node, as already discussed in
Ref.~\cite{b1}.  Let us consider a network which is built by attaching every
new node to exactly $p$ earlier nodes.  For the linear attachment kernel, the
degree distribution $N_k(t)$, which is now defined only for $k\geq p$,
evolves according to
\begin{equation}
\label{Nkp}
{d N_k\over dt}={p\over M_1}
\left[(k-1) N_{k-1}-k N_k\right]+\delta_{kp}.
\end{equation}
Clearly, the average in-degree $\langle i\rangle$ and out-degree $\langle
j\rangle$ of each node in this network is equal to $p$.  By applying the
basic approach of Sec.~III to this rate equation, we find that the degree
distribution again asymptotically approaches a stable distribution $N_k\to
tn_k$, with
\begin{equation}
\label{nkp}
n_k={2p(p+1)\over k(k+1)(k+2)} \qquad{\rm for}\quad k\geq p. 
\end{equation}
Thus for the linear attachment kernel, the average node degree does not
affect the exponent $\nu$ of the degree distribution.  However, for other
solvable examples, the new feature of attaching the new node to more than one
pre-existing node leads to different degree distributions.  For example, for
the shifted linear kernel we find
\begin{eqnarray}
\label{nkwp}
n_k&=&{\rm const.}\times 
{\Gamma(k+w)\over \Gamma(k+3+w+w/p)} \quad{\rm for\ } k\geq p,\\
n_p&=&\left(1+p\,{p+w\over
    2p+w}\right)^{-1}.  
\label{np}
\end{eqnarray}
This gives the asymptotic behavior $n_k\sim k^{-(3+w/p)}$.  Thus the exponent
of the degree distribution {\em depends} on the average node degree, with
$\nu=3+w/p$.

The multiple linking construction also reduces the number of nodes with
in-degree zero.  For example, for the GN with the shifted linear attachment
kernel, the fraction of such nodes is $n_1=(2+w)/(3+2w)$, which is always
larger than 1/2.  However, for the multiple linking construction, the
fraction of nodes with in-degree zero is reduced to the value $n_p$ given in
Eq.~(\ref{np}).  If we use $p=7$ to reproduce the correct average node degree
of the web graph, the fraction of nodes with in-degree zero always exceeds
1/8, which, however, apparently disagrees with web data \cite{brod}.  Thus
while multiple attachment does reduce the number of poorly-connected nodes,
this reduction is still insufficient to account for web-graph data.  However,
it is clear that the multiple linking construction has the potential to
provide a better description of citation data.

Another shortcoming of the multiple attachment construction is that it cannot
dynamically generate a non-trivial out-degree distribution.  However, we can
extend the GN model by allowing for creation of links between existing
nodes\cite{krr}.  This simple construction allows us to generate non-trivial
out-degree distributions which closely match web graph data.  An even more
challenging direction is to describe the global topological structure of
growing networks.  The GN model leads to a single-component tree graph, while
the web graph has numerous disconnected components.  A deeper understanding
of the web graph may provide valuable insights to help develop algorithms for
web crawling, searching, and community discovery.

We are grateful to NSF grant DMR9978902 and ARO grant DAAD19-99-1-0173 for
partial financial support of this research.

\end{multicols}

\begin{thebibliography}{99}
\bibitem{note1}
  The apparently singular behavior at $y=2$ indicates that a scale finer than
  linear in $k$ is required in the proximity of $y=2$.  The proper scale is
  $\sqrt{k}$ and its use resolves the singularity and matches solutions for
  $y<2$ and $y>2$.
\end{thebibliography}
\end{document}